\documentclass[conference,a4paper]{IEEEtran}
\IEEEoverridecommandlockouts
\usepackage{cite}
\usepackage{makecell}
\usepackage[utf8x]{inputenc}
\usepackage{graphicx}
\usepackage{array}
\usepackage{balance}
\usepackage{booktabs}
\usepackage{url}
\usepackage[misc]{ifsym}
\usepackage[dvips]{color}
\usepackage{mathrsfs}
\usepackage[ruled,linesnumbered]{algorithm2e}
\usepackage{amsmath}
\usepackage{amsthm}
\usepackage{amsfonts,amssymb}
\usepackage{multirow}
\usepackage{subfigure}

\begin{document}
	  
\title{Soft Multipath Information-Based UWB Tracking in Cluttered Scenarios: Preliminaries and Validations}

\author{\IEEEauthorblockN{
Chenglong~Li\textsuperscript{1,2}$^\star$,
Zukun~Lu\textsuperscript{1,2},
Long~Huang\textsuperscript{1,2},
Shaojie~Ni\textsuperscript{1,2},
Guangfu~Sun\textsuperscript{1,2},
Emmeric~Tanghe\textsuperscript{3},   
Wout~Joseph\textsuperscript{3}
} 
\IEEEauthorblockA{\textsuperscript{1}
College of Electronic Science and Technology, National University of Defense Technology, 410073 Changsha, China}
\IEEEauthorblockA{\textsuperscript{2}
Key Laboratory of Satellite Navigation Technology, 410073 Changsha, China}
\IEEEauthorblockA{\textsuperscript{3}
WAVES group, Department of Information Technology, Ghent University-imec, 9052 Ghent, Belgium} 
 \IEEEauthorblockA{$\star$ Email: \emph{chenglong.li@nudt.edu.cn} }
}

\maketitle

\begin{abstract}
In this paper, we investigate ultra-wideband (UWB) localization and tracking in cluttered environments. Instead of mitigating the multipath, we exploit the specular reflections to enhance the localizability and improve the positioning accuracy. With the assistance of the multipath, it is also possible to achieve localization purposes using fewer anchors or when the line-of-sight propagations are blocked. Rather than using single-value distance, angle, or Doppler estimates for the localization, we model the likelihoods of both the line-of-sight and specular multipath components, namely soft multipath information, and propose the multipath-assisted probabilistic UWB tracking algorithm. Experimental results in a cluttered industrial scenario show that the proposed algorithm achieves 46.4~cm and 33.1~cm 90th percentile errors in the cases of 3 and 4 anchors, respectively, which outperforms conventional methods with more than 61.8\% improvement given fewer anchors and strong multipath effect. 

\end{abstract}


\section{Introduction}
Accurate indoor location awareness can help to automate and optimize the flows of many vertical applications, such as autonomous ground or aerial vehicle, smart logistics, etc. Conventional localization approaches exploit the received signal strength indicator (RSSI) for localization purposes. However, the RSSI is a representation of the signal power level, which is easily affected by the channel fading. In recent years, there are an increasing number of commodity radio devices supporting channel response measurement at a relatively low cost. This attracts extensive industrial and academic attention, as this channel information allows the localization algorithm design from the perspective of the physical layer \cite{SpotFi2015,Ma2020}. Many channel response-based methods have been proposed to solve the localization and tracking challenges in cluttered environments. Herein, we give an overview of the state-of-the-art (SOTA) methods with a focus on the solutions in the case of multipath-rich scenarios. The SOTA solutions are primarily classified into three categories: multipath-separation \cite{Sen2013,Kuhn2010,Li2022_mamimo}, multipath-compensation \cite{Wymeersch2012,Kegen2019}, and multipath-assistance approaches \cite{Witrisal2016,Leitinger2019,Chi2021,Wen2021}, which are elaborated below.
\par
\textit{Multipath-separation scheme}: Within this scheme, multiple antennas or wideband is required. Take the multi-antenna system as an example, the angle-of-arrival or angle-of-departure of multipath components (MPCs) can be estimated via a super-resolution channel estimation algorithm. Then the angle estimate of the direct link is separated from the other MPCs and used for localization and tracking \cite{Sen2013}. For a wideband system, a similar procedure can be implemented but using the time-of-flight (ToF) or range estimate of the direct link.
\par
\textit{Multipath-compensation scheme}: In practice, the resolution of the angle or distance estimate is constrained by lots of factors \cite{Li2022_mamimo}, such as the number of antennas, bandwidth, the accuracy of the adopted super-resolution channel parameter estimation algorithms, etc. We cannot always expect a satisfying direct link separation from multipath propagation. Besides, in some cases, the line-of-sight (LoS) links are blocked resulting in no LoS estimate at all. To this end, there are other prevalent solutions that use machine learning tools to calibrate the angle or distance estimates \cite{Wymeersch2012,Kegen2019,Yanru2022}.
\par
\textit{Multipath-assistance scheme}: There are increasing solutions exploiting multipath propagation for localization, as it can provide a richer geometrical diversity compared with using the LoS merely. There are generally two prevalent solutions. One is the fingerprinting-based approach \cite{Kram2019,Chi2021}, which does not obtain the MPCs directly but trains a regression or classification model from a labelled dataset. The machine learning tools take care of multipath information extraction implicitly. The other popular scheme is to first estimate the MPCs using a super-resolution channel estimation algorithm, then associate the MPCs to the corresponding virtual anchors. Obtaining the locations of the virtual anchors generally requires prior knowledge of the floor plan \cite{Witrisal2016}. Alternatively, it is within the framework of simultaneous localization and mapping \cite{Leitinger2019}, which obtains the locations of the users and the reflectors at the same time.
\par
Considering the ultra-wideband (UWB)-based tracking in cluttered industrial environment, we propose a soft multipath information-assisted tracking algorithm. Instead of estimating the deterministic ranges of MPCs, we model the ranging likelihood of the specular MPCs based on a sequential spatial-temporal-spatial likelihood mapping.

\section{Signal Model and Probabilistic Localization}
\label{sec:MALoc}
\subsection{Signal Model}
Consider a scenario as shown in Fig.~\ref{fig:LocScenario}(a), there are $N$ UWB access points (or anchors) with known locations ${\bf{p}}_{\rm{A}}^{(i)}, i=1,\cdots,N$. The tag at position ${\bf{p}}_{\rm{T}}$ is the user to be located. Generally, to localize the user with respect to UWB devices, the distance between the anchor ${\bf{p}}_{\rm{A}}^{(i)}$ and the tag ${\bf{p}}_{\rm{T}}$ should be inferred via the channel measurement, specifically CIR. The CIR $h^{(i)}(\tau)$ between ${\bf{p}}_{\rm{A}}^{(i)}$ and ${\bf{p}}_{\rm{T}}$ in equivalent baseband notation is given by 
\begin{equation}\label{eq:channel}
\!\!h^{(i)}(\tau)\!=\underbrace{\alpha^{(i)}_0\delta(\tau-\tau^{(i)}_0)\!+\!\sum_{l=1}^{L_i}\alpha^{(i)}_l\delta(\tau-\tau^{(i)}_l)}_{\rm{SMCs}}\!+\!\underbrace{\nu^{(i)}(\tau)}_{\rm{stochastic}},\!
\end{equation}where the channel is modeled as the superposition of two parts, namely the specular multipath components (SMCs) and the stochastic part including dense multipath components (DMCs), diffuse scattering, and measurement noise \cite{Tanghe2014}. In the first term of \eqref{eq:channel}, $\alpha_l$ and $\tau_l$ denote the complex amplitude and delay of the deterministic MPCs, respectively, and $L_i$ is the number of specular reflections of the $i$-th anchor. $\alpha_0$ and $\tau_0$ define the LoS component, whereas $\alpha_0=0$ indicates no LoS component. The second term of \eqref{eq:channel} describes the second-order statistics of the radio channel via a zero-mean Gaussian random process.

\subsection{Probabilistic Localization}
In conventional multipath-separation methods, the LoS ToFs $\tau^{(i)}_0, i=1,\cdots,N$ of the $N$ anchors are estimated. Then, the location of the tag can be calculated via trilateral positioning. However, the accuracy of range estimation degrades distinctly in the case of multipath propagation and non-line-of-sight (NLOS) conditions. Many efforts have been devoted to improving the accuracy, such as adopting machine learning-based ranging correction \cite{Wymeersch2012,Kegen2019,Yanru2022}. Alternatively, the idea of ranging likelihood or soft range information was proposed, which uses not only the most likely distance but also other probable ranging estimates with probabilities, rather than a single-value range estimate, for more accurate positioning \cite{Mazuelas2018,Conti2019}. Define $y_1,y_2,\cdots,y_N$ the range-related measurement, e.g., ToF in the UWB systems. From the perspective of Bayesian estimation, the tag's location can be inferred via maximizing the posterior distribution of ${\bf{p}}_{\rm{T}}$ given the range-related measurement,
\begin{equation}\label{eq:LoS_Loc}
\begin{aligned}
\!\!\hat{{\bf{p}}}_{\rm{T}}&=\!\underset{{\bf{p}}_{\rm{T}}}{\arg\max} f({\bf{p}}_{\rm{T}}|y_1,\cdots,y_N), \\   
&=\!\underset{{\bf{p}}_{\rm{T}}}{\arg\max}f\!\left(y_N{\large|}\left\|{\bf{p}}_{\rm{T}}-\!{\bf{p}}_{\rm{A}}^{(N)}\right\|\right)\!f({\bf{p}}_{\rm{T}}|y_1,\cdots,y_{N-1}),\!\!\\
&=\!\underset{{\bf{p}}_{\rm{T}}}{\arg\max} f({\bf{p}}_{\rm{T}})\prod_{i=1}^{N}f\left(y_i{\large|}\left\|{\bf{p}}_{\rm{T}}-{\bf{p}}_{\rm{A}}^{(i)}\right\|\right),
\end{aligned}
\end{equation}where $f({\bf{p}}_{\rm{T}})$ is the prior distribution of ${\bf{p}}_{\rm{T}}$. $f(y_i|\|{\bf{p}}_{\rm{T}}-{\bf{p}}_{\rm{A}}^{(i)}\|)$ is the distribution of range-related measurement $y_i$ conditioned on the distance $\|{\bf{p}}_{\rm{T}}-{\bf{p}}_{\rm{A}}^{(i)}\|$. The ranging likelihood $\mathcal{L}_{y_i}(\|{\bf{p}}_{\rm{T}}-{\bf{p}}_{\rm{A}}^{(i)}\|)$ is proportional to $f(y_i|\|{\bf{p}}_{\rm{T}}-{\bf{p}}_{\rm{A}}^{(i)}\|)$ \cite{Mazuelas2018}.
\par

\begin{figure}[t]
\centering
\setlength{\abovecaptionskip}{-0.15cm}
\setlength{\belowcaptionskip}{-0.15cm}
\includegraphics[width=0.49\textwidth]{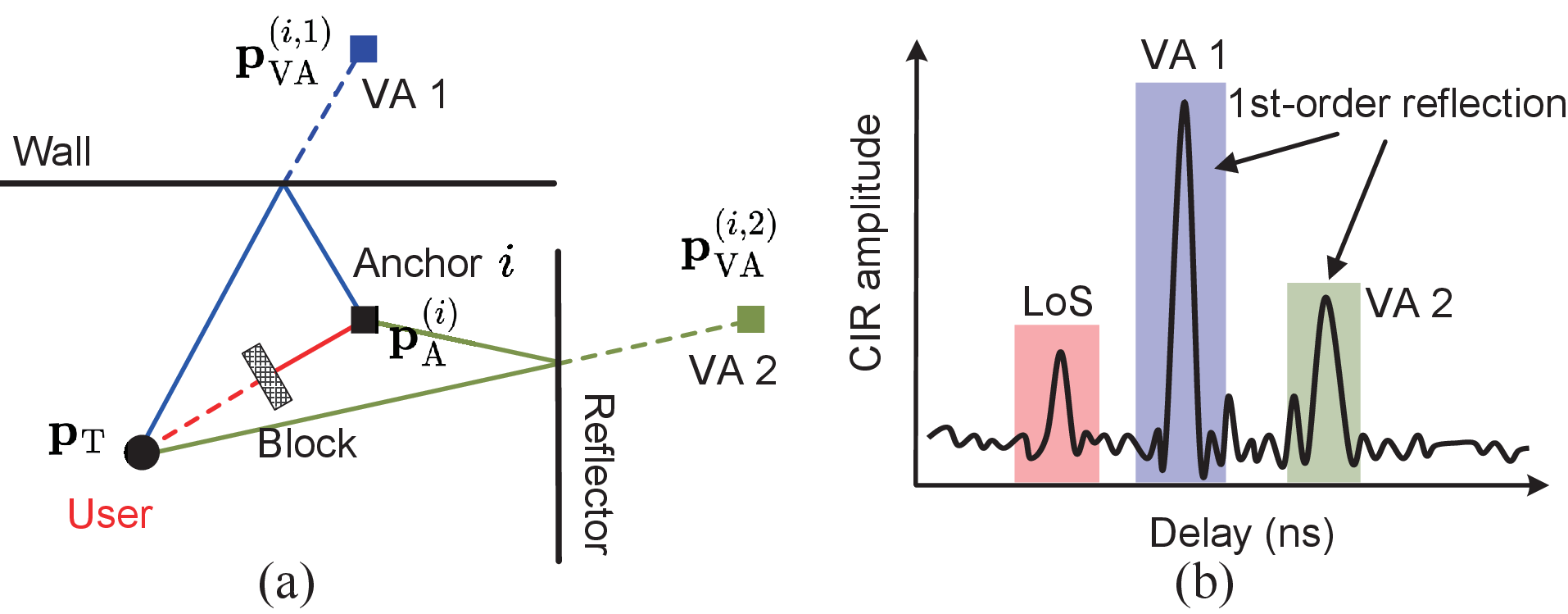}
\caption{Illustration of (a) multipath-assisted localization and (b) the corresponding CIR for the $i$-th UWB anchor.}
\label{fig:LocScenario}
\end{figure}

However, the above ranging likelihood considered the LoS distance merely. How to encapsulate the MPCs, in the framework of probabilistic localization, to enhance locatability using fewer anchors is an interesting topic. As shown in Fig.~\ref{fig:LocScenario}(a), the specular reflections can be regarded as the signal transmitted by the virtual anchors that are at the mirrored locations of the physical anchors with respect to (w.r.t.) the wall or other planar reflectors. Then, the location estimation in \eqref{eq:LoS_Loc} for the case considering the SMCs can be rewritten as,
\begin{equation}\label{eq:MPC_Loc}
\begin{aligned}
\hat{{\bf{p}}}_{\rm{T}}=\underset{{\bf{p}}_{\rm{T}}}{\arg\max}f&({\bf{p}}_{\rm{T}})\prod_{i=1}^{N}f\left(y_i{\large{|}}\left\|{\bf{p}}_{\rm{T}}-{\bf{p}}_{\rm{A}}^{(i)}\right\|\right)\\
&\quad\times\prod_{i=1}^{N}\prod_{l=1}^{L_i}f\left(y_{i,l}^{\rm{VA}}{\large{|}}\left\|{\bf{p}}_{\rm{T}}-{\bf{p}}^{(i,l)}_{\rm{VA}}\right\|\right),
\end{aligned}
\end{equation}where $y_{i,l}^{\rm{VA}}$ represents the range-related measurement corresponding to the $l$-th virtual position ${\bf{p}}^{(i,l)}_{\rm{VA}}$ of the $i$-th physical anchor. To the authors' best knowledge, it is the first time to propose multipath-assisted probabilistic localization, which utilizes the soft information of both LoS and SMCs. Note that \eqref{eq:MPC_Loc} is under the assumption that the SMCs are unrelated. Using the MPCs for localization is built upon the idea of improving the geometrical diversity based on propagation from different directions. If the SMCs are related, it means the physical anchor and the corresponding virtual anchors are very likely spatially closed. In this case, the geometrical diversity is not sufficient and thus the enhancement by the multipath is limited.

\section{Soft Multipath Information-Assisted Tracking: Model and Practical Implementation}
\label{sec:algDesign}
With the maximum a posterior estimator in \eqref{eq:MPC_Loc}, we can infer the user's location. The challenge therein will be how to obtain the conditional distribution of the range-related measurement which is proportional to the ranging likelihood. In available solutions, the ranging likelihood is estimated from channel measurements using the learning-based scheme. Specifically, the channel information and the corresponding real distances are used to train the generative model \cite{Mazuelas2018,Yuxiao2021}, namely the conditional distribution of the range-related measurement. During the online phase, the new channel measurement is fed to the generative model to obtain the ranging likelihood. However, the training of the generative model is challenging, as this requires a massive labeled dataset and expensive training resources. Furthermore, available learning-based schemes train the ranging likelihood for the LoS and NLoS cases separately, which requires an extra LoS identifier.

\begin{figure}[t]
\centering
\setlength{\abovecaptionskip}{-0.05cm}
\setlength{\belowcaptionskip}{-0.1cm}
\includegraphics[width=0.39\textwidth]{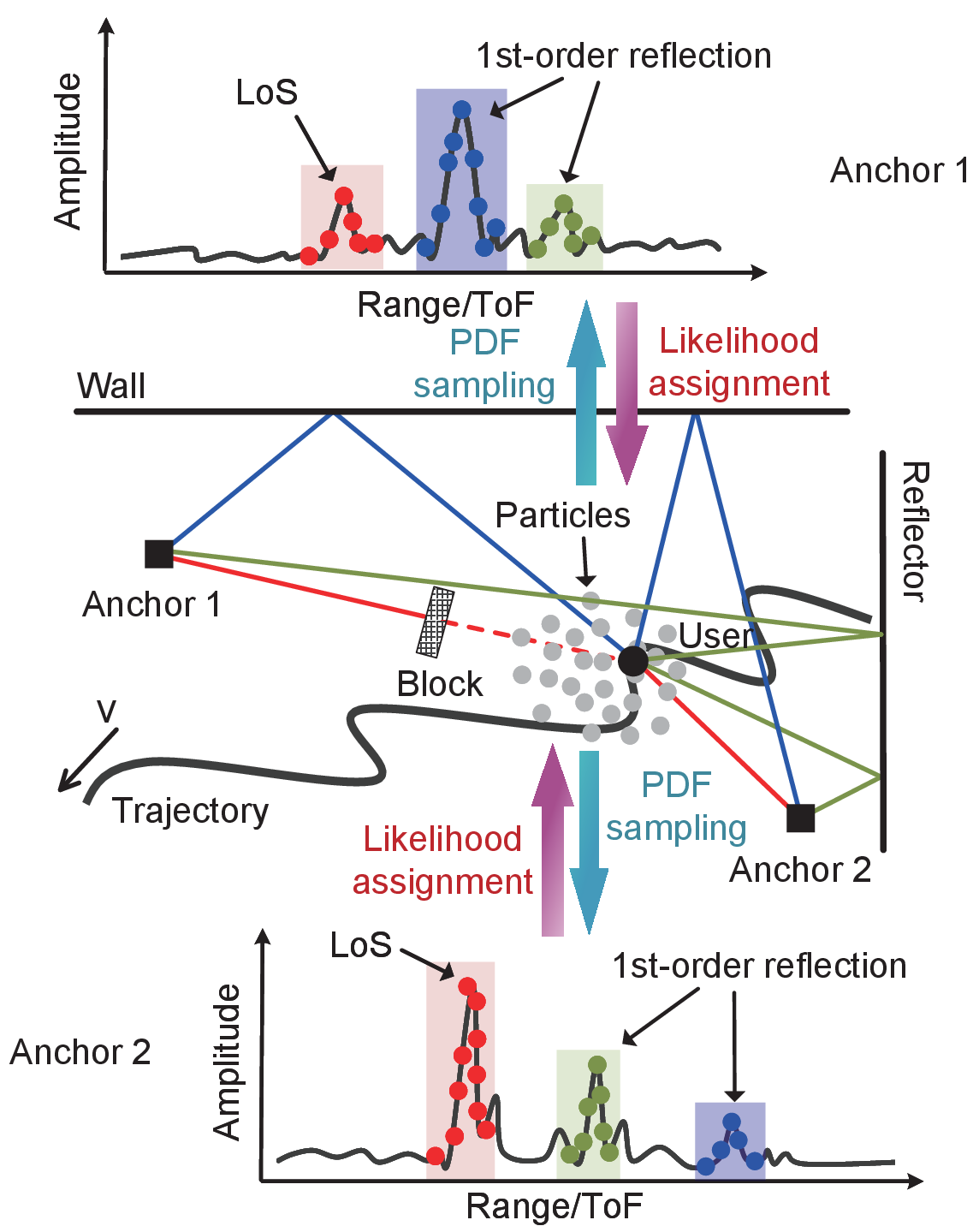}
\caption{Illustration of soft multipath information-assisted UWB tracking via the spatial-temporal-spatial likelihood mapping.}
\label{fig:MPCsMapping}
\end{figure}
\par
Suppose we have CIR measurements that are reported by the UWB radio devices, we can regard the CIR measurement $\vert h(\tau)\vert$ as the amplitude-delay profile as shown in Fig.~\ref{fig:LocScenario}(b). Suppose we have the channel frequency response (CFR) measurement, then an inverse Fourier transfer (IFT) can be conducted for the wideband system to obtain the amplitude-delay profile. The peaks of the amplitude-delay profile represent the strength of MPCs. In other words, they also roughly indicate the corresponding possibility of specific delays. However, if we want to use the MPCs for localization, the likelihood of each MPC is required as shown in Fig.~\ref{fig:LocScenario}(b). Note that herein we only consider the LoS component and the first-order specular reflections as the signals from the second or higher-order bounces are weak. 
\par
An intuitive method to obtain the likelihood at a specific component is to truncate the amplitude-delay profile, as shown in Fig.~\ref{fig:LocScenario}(b). Namely, the truncated profile at the vicinity of the peak is regarded as the corresponding ranging likelihood. We normalize the amplitude-delay profile via dividing it by the corresponding integral in the delay domain. Then, the probability density function (PDF) of the SMC at $i$-th anchor is approximated as follows,
\begin{equation}\label{eq:smcPDF}
\mathcal{L}_{i}(\tau_l^{(i)})\propto \frac{h^{(i)}(\vert\tau-\tau_l^{(i)}\vert<\epsilon)}{\int_{\tau_l^{(i)}-\epsilon}^{\tau_l^{(i)}+\epsilon}h^{(i)}(\vert\tau-\tau_l^{(i)}\vert<\epsilon)\rm{d}\tau},
\end{equation}where $\epsilon$ denotes the radius of neighborhood at delay $\tau_l^{(i)}$.
\par
However, how to accurately identify the SMC peaks and the corresponding neighborhood radius is challenging. To obtain the ranging likelihood of each SMC, as shown in Fig.~\ref{fig:MPCsMapping}, we propose a spatial-temporal-spatial likelihood mapping algorithm built upon the idea of sequential importance sampling in the particle filter. The key idea of importance sampling is to approximate the posterior PDF $p(\textbf{x}|\textbf{z})$ via a batch of samples $\textbf{x}_k, (k=1,\cdots, K)$ and the associated weights $\textbf{w}_k, (k=1,\cdots, K)$, that are generated from a proposal distribution (or an importance density), e.g., a multivariate Gaussian distribution. Then, the posterior PDF can be approximated as \cite{Arulampalam2002},
\begin{equation}\label{eq:SIS}
p(\textbf{x}|\textbf{z})=\sum_{k=1}^{K}\hat{\textbf{w}}_k\cdot\delta(\textbf{x}-\textbf{x}_k),\hspace{0.2cm} {\hat{\textbf{w}}_k}=\frac{{\textbf{w}}_k}{\sum_{k=1}^{K}{\textbf{w}}_k},
\end{equation}where weight $\textbf{w}_k$ can be calculated from the measurement density and the proposal density \cite{Arulampalam2002}. It is interesting to observe that \eqref{eq:SIS} has a similar expression as the CIR in baseband notation in \eqref{eq:channel}. The weight can also be regarded as the normalized amplitude of each MPC. This means we can use the discretized CIR to approximate the ranging likelihood of each SMC given the sufficient proposal locations. Specifically, we generate a batch of particles based on the motion model and use the distance calculated from the particles to sample the CIR of each pair of UWB anchor-tag link, namely \textit{spatial-temporal mapping} for PDF sampling, indicated by the light blue arrows in Fig.~\ref{fig:MPCsMapping}. After this, we assign the ranging likelihood of all pairs of links to the particle, represented by \textit{temporal-spatial mapping} for location likelihood assignment, indicated by the purple arrows in Fig.~\ref{fig:MPCsMapping}. All these procedures are compatible with the particle filter, so we can include the location likelihood approximation into the framework of particle filter for tracking purposes. 
\par
The specific procedures of the algorithm are given below. 
\par
\begin{description}
\item[\textbf{S-1}] Generate $K$ particles via the state transition model, which is assumed as a constant velocity model for human or vehicle tracking, given by,
\begin{equation}\label{eq:MotionModel}
\begin{bmatrix}
\textbf{P}^{\top}_{(t+1)}\\ 
\textbf{v}^{\top}_{(t+1)}
\end{bmatrix}=\begin{bmatrix}
\mathbf{I}_{2} & \Delta t\cdot\mathbf{I}_{2}\\ 
\mathbf{0} & \mathbf{I}_{2}
\end{bmatrix}\begin{bmatrix}
\textbf{P}^{\top}_{(t)}\\ 
\textbf{v}^{\top}_{(t)}
\end{bmatrix}+\begin{bmatrix}
\Delta t\cdot\mathbf{1}_{2\times1}\\ 
\mathbf{1}_{2\times1}
\end{bmatrix}n_v,
\end{equation}where $\textbf{P}$ and $\mathbf{v}$ are the coordinates and velocity of the moving tag, respectively. For the sake of simplicity, only two-dimensional (2-D) location coordinates and velocity are considered. $\Delta t$ is the elapsed time between two consecutive timestamps $t+1$ and $t$. ${n}_v$ is the zero-mean Gaussian velocity error. To initialize the particle filter for tracking, the first location of the user is needed, which can be estimated by, e.g., the conventional multipath separation-based algorithm.
\item[\textbf{S-2}] Calculate the distances or delays from the $K$ particles to the physical and virtual anchors. The locations of the physical anchors and some planar reflectors (e.g., walls, metal plates) are \textit{a priori} knowledge.
\item[\textbf{S-3}] Use the calculated delays to resample the amplitude-delay profile $\vert h(\tau)\vert$, as indicated in Fig.~\ref{fig:LocScenario}(b). Normalize the truncated amplitude-delay profile via \eqref{eq:smcPDF} and obtain the marginal ranging PDF of each SMC. In Fig.~\ref{fig:MPCsMapping}, dots with different colors represent the resampled marginal profiles for different SMCs.
\item[\textbf{S-4}] Assign the delay likelihood to each particle and combine the corresponding likelihood from different anchors of the same particle, defined as $\mathcal{L}(\mathbf{P}_k)$, $k=1,\cdots,K$. In this way, we obtain the location likelihood of each particle, namely the equivalent likelihood function in \eqref{eq:MPC_Loc}.
\item[\textbf{S-5}] Normalize the location likelihood for the $K$ particles (e.g., 200), to $(0,1]$. The $k$-th particle for updating is weighted via $\hat{\omega}_{k}$, defined as
\begin{equation}\label{eq:PF_weight}
\ln\hat{\omega}_{k}={\frac{1}{2\sigma_c^2}\left(\min\limits_{1\leq k\leq K}\left\lbrace{\omega_{k}}\right\rbrace-\omega_{k}\right)},
\end{equation}where $\omega_{k}=\left(1-\mathcal{L}(\mathbf{P}_k)\right)^2$. $\sigma_c$ is the standard deviation of $\mathcal{L}(\mathbf{P})$. To avoid the degeneracy phenomenon in PF, systematic resampling is adopted \cite{Arulampalam2002}. When receiving new CIR measurements, repeat \textbf{S-1} to \textbf{S-5}.
\end{description}

\par
In this paper, we focus on the CIR for range or delay domain processing. However, the proposed idea can be analogously generalized to other domain features, such as angle, Doppler velocity, etc., for localization and tracking. When we have the channel frequency response for a multi-antenna system, for example, a Fourier transfer can be conducted to obtain the amplitude-angle profile, which can be approximated as the angular likelihood. Besides, we can also adopt, e.g., MUltiple Signal Identification and Classification algorithm, to obtain the delay or angle spectrum. Then sequential importance sampling is used to approximate the ranging or angular likelihood.

\section{Experimental Evaluation}
\label{sec:expResults}

\subsection{Experiment and Dataset}
\label{sec:MeasScenario}

\begin{figure}[t]
\centering
\setlength{\abovecaptionskip}{-0.1cm}
\setlength{\belowcaptionskip}{-0.1cm}
\includegraphics[width=0.49\textwidth]{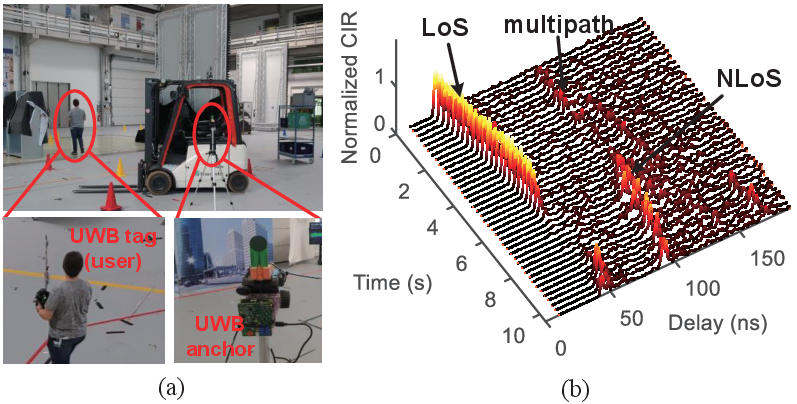}
\caption{(a) The measurement campaign in a typical industrial environment. (b) Example of CIR measurements indicating multipath and NLoS propagations.}
\label{fig:MeasCamp}
\end{figure}

The proposed algorithm has been evaluated via a real-world collected UWB CIR dataset, which is open-access and elaborated in \cite{IPIN2021}. The data (\textit{i.e.}, CIR) is recorded based on Decawave DW1000 chipset with a center frequency of 3993.6~MHz and bandwidth 499.2~MHz. The receivers (anchors) are placed at a height of 1.5~m. The transmitter (tag) is carried by a participant, as shown in Fig.~\ref{fig:MeasCamp}, of which the ground truth is obtained via a millimeter-accurate Qualisys motion capture system. UWB signals is regularly transmitted and received by the UWB anchors. The experimental environment mimics an industrial warehouse of an area of about 300~m$^2$ with a partial enclosure by reflecting walls and some arbitrarily deployed metallic objects, e.g., industrial vehicles or metal shelves, which are typical for industrial scenarios. For the floor plan, the locations of two large reflectors including one wall and one metallic reflector are available, which can be measured when obtaining the locations of UWB anchors with some extra efforts. Due to page limitation, we omit the details of the measurement setting here but refer to \cite{IPIN2021}.

\subsection{Results and Analysis}
In this subsection, we evaluate the performance of the proposed soft multipath information-assisted tracking algorithm. We set the localization results using leading edge detection (LDE) based ToF estimation \cite{Kuhn2010} as the benchmark. The LDE is used to detect the leading edge of the first peak, which separates the LoS from the other MPCs and only uses the LoS range estimation for localization. For the tracking filter, we adopt a particle filter in which the parameter settings are the same as the proposed algorithm with 1000 particles. 
\par
\begin{figure}[t]
\centering
\setlength{\abovecaptionskip}{-0.1cm}
\setlength{\belowcaptionskip}{-0.1cm}
\includegraphics[width=0.495\textwidth]{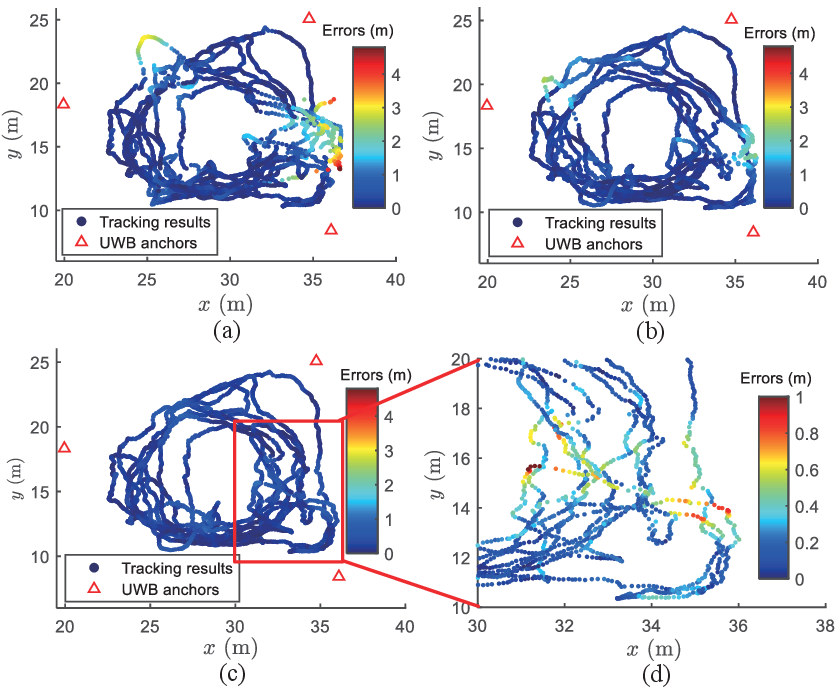}
\caption{Tracking results with the errors along the trajectory in the case of three UWB anchors (indicated by red triangles): (a) Benchmark. (b) Proposed method with LoS merely. (c) Proposed method with SMC. (d) Partial enlargement of (c).}
\label{fig:three_anchors_traj}
\end{figure}

Fig.~\ref{fig:three_anchors_traj} shows the tracking results in the case of three anchors. The color depth of the tracking results along the trajectory indicates the location errors. As shown in Fig.~\ref{fig:three_anchors_traj}(a), the ToF estimation-based method (\textit{i.e.}, benchmark) has poor tracking results with errors up to 5~m, especially when one or more UWB anchors is blocked by the obstacles. In this case, the range estimation errors are large, which greatly degrades the localization performance when only the number of anchors is limited. Fig.~\ref{fig:three_anchors_traj}(b) shows the tracking results using the proposed algorithm considering the likelihood of LoS merely, whereas Fig.~\ref{fig:three_anchors_traj}(c) presents the tracking results of the proposed algorithm with SMCs. We can observe that the proposed method can mitigate the tracking errors distinctly (the maximal error is about one meter) in the case of multipath.
\par

\begin{figure}[t]
\centering
\setlength{\abovecaptionskip}{-0.1cm}
\setlength{\belowcaptionskip}{-0.1cm}
\includegraphics[width=0.47\textwidth]{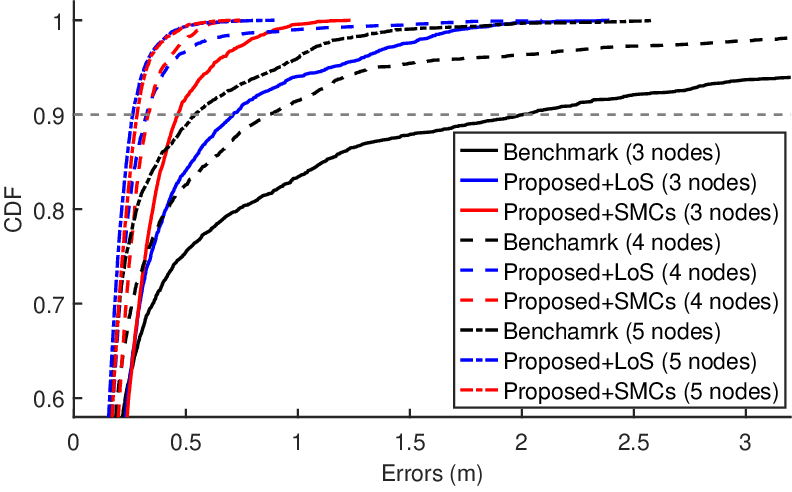}
\caption{CDF of the tracking errors regarding three and four UWB anchors.}
\label{fig:3_4_5anchors_CDF}
\end{figure}

We summarize the tracking performance via the cumulative distribution function (CDF) of tracking errors in Fig.~\ref{fig:3_4_5anchors_CDF}. When we only concern the median errors, the performance of these three solutions is similar with differences of less than 4 cm. However, when we compare the 90-th percentile and 95-th percentile tracking errors, the proposed probabilistic tracking algorithm shows its superiority. Specifically, when we compare the performance of the proposed method with benchmark in the case of three anchors, the proposed method can mitigate the outliers effectively. It has achieved 64.5\% accuracy improvement given the 90-th percentile errors (from 200~cm to 71.1~cm). Further, when consider the performance of the proposed method with SMCs, 90-th percentile errors are 46.4~cm and 33.1~cm regarding three and four anchors, which greatly outperform the benchmark with 200~cm and 86.6~cm with 76.8\% and 61.8\% improvement, respectively. 
\par
Furthermore, in case that five UWB anchors are adopted, the tracking performance of the proposed method with LoS approaches the performance of the proposed method with SMCs, except for limited outliers. These observations indicate that the proposed probabilistic algorithm is more robust than the deterministic method via ToF estimation, especially in a cluttered scenario. It is also beneficial to the cases that only a limited number of anchors are available or there are sufficient anchors but most of them are under NLoS condition.

\section{Conclusion}
\label{sec:conclusion}
In this paper, we have investigated the UWB tracking problem in a cluttered indoor industrial environment. Instead of mitigating the multipath, we propose to utilize the soft multipath information for localization. An efficient implementation built upon the sequential importance sampling has presented and analysed. According to the experimental validation, the proposed algorithm performs well in cluttered scenarios even if a limited number of UWB anchors are adopted. Future work will consist of incorporating the probabilistic localization of reflectors or virtual anchors in the framework of multipath-assisted tracking via simultaneous localization and mapping. 
\section*{Acknowledgment}
This work is supported in part by the National Key R\&D Program of China under Grant 2023YFC2205400, by the National Natural Science Foundation of China under Grant U20A0193 and 62303482, and by the imec project UWB-IR AAA. The authors would like to thank Sebastian Kram, Maximilian Stahlke, and Christopher Mutschler from Fraunhofer Institute for Integrated Circuits IIS for the measurement.


\bibliographystyle{IEEEtran}
\bibliography{MapTrack}

\end{document}